# Nanoimprint of a 3D structure on an optical fiber for light wavefront manipulation

Giuseppe Calafiore[†,1], Alexander Koshelev[†,1], Frances I. Allen[2,3], Scott Dhuey[3], Simone Sassolini[3], Edward Wong[3], Paul Lum[2], Keiko Munechika[1,*], Stefano Cabrini[3].

[1]aBeam Technologies Inc., 22290 Foothill Blvd. St. 2, Hayward, CA 94541
[2]Biomolecular Nanotechnology Center/QB3, Stanley Hall, University of California, Berkeley, CA 94720
[3]Molecular Foundry, Lawrence Berkeley National Lab, 67 Cyclotron Rd, Berkeley, CA 94720

[†]G. Calafiore and A. Koshelev contributed equally to this work
*E-mail: km@abeamtech.com

**Abstract**
Integration of complex photonic structures onto optical fiber facets enables powerful platforms with unprecedented optical functionalities. Conventional nanofabrication technologies, however, do not permit viable integration of complex photonic devices onto optical fibers owing to their low throughput and high cost. In this paper we report the fabrication of a three dimensional structure achieved by direct Nanoimprint Lithography on the facet of an optical fiber. Nanoimprint processes and tools were specifically developed to enable a high lithographic accuracy and coaxial alignment of the optical device with respect to the fiber core. To demonstrate the capability of this new approach, a 3D beam splitter has been designed, imprinted and optically characterized. Scanning electron microscopy and optical measurements confirmed the excellent lithographic capabilities of the proposed approach as well as the desired optical performance of the imprinted structure. The inexpensive solution presented here should enable advancements in areas such as integrated optics and sensing, achieving enhanced portability and versatility of fiber optic components.

**Introduction**

Optical fibers have found many applications both in industry and in academic research labs due to their ease of use and integration[1]. In recent times, new types of fiber optic devices have emerged in which photonic structures are fabricated directly on a fiber facet. Examples of this are lab-on-a-fiber[2], localized surface plasmon biosensors[3,4], an axicon lens[5,] opto-mechanical sensors[6] and vortex beam generators[7]. These devices are potentially transformative in fields like integrated optics, biosensors and medical probes. In addition, this form of direct integration offers advantages in terms of versatility, portability and flexibility of the miniaturized optical components, which can also be fabricated on fiber-coupled devices such as lasers, detectors, splitters and circulators. Three-dimensionality is especially important for powerful photonic functionalities, because it enables engineering of the relative phase shift of light and manipulation of its wavefront[5,7]. Nanofabrication of 3D structures on a surface as small as the end of an optical fiber remains the main challenge and bottleneck of this integration. Since most of the fabrication capabilities are designed for wafer-scale manufacturing, photonic structures on a fiber have been fabricated mainly using time-consuming and expensive techniques such as electron beam lithography (EBL)[2,3] and focused ion beam (FIB)[5], which dramatically limit the availability of these devices. Two-photon lithography can produce very complex 3D geometries, but it has a low throughput and limited resolution[8].

In this paper we propose the fabrication of 3D (multilevel) structures for light wavefront manipulation on the facet of an optical fiber by Ultraviolet Nanoimprint Lithography (UV-NIL). NIL is a mechanical lithography technique capable of high resolution and throughput at a low cost[9,10]. Low manufacturing costs are particularly important for the fabrication of disposable fiber optic probe sensors. NIL has been used to pattern 1D metal gratings as well as biomimetic patterns on an optical fiber[11,12]. However, the ability to imprint 3D structures with good lithographic fidelity as well as accurate alignment between the fiber mode and the photonic device is required in many applications[5,7]. Here, we exploit the unique capability of NIL to replicate 3D geometries in one step using a functional material. The choice of a transparent resist as both the imprint polymer and the diffractive medium permits this single-step fabrication. To demonstrate manipulation of the wavefront of light using such a device, we designed a wavelength selective 3D beam splitter. The design, fabrication and characterization of the fiber optic splitter is reported in detail below. The processes developed for the fabrication of this structure can be implemented for the fabrication of a multitude of other optical components, paving the way to a new class of inexpensive fiber optic integrated devices and probes.

**Fabrication**

In order to maintain lithographic fidelity, the processes and tools typically used to imprint 3D structures on planar substrates must be appropriately modified to enable their application to imprint on an optical fiber. A detailed overview of the proposed fabrication approach is reported below and illustrated in Fig. 1. The imprint master mold with intruded, multilevel features is obtained by grayscale Gallium-FIB milling on a flat substrate (Fig. 1a). Ideally, the mold should be transparent to allow optical inspection and alignment between the fiber and the mold at the imprint stage, thus milling should be performed on quartz or another optically transparent material. However, in this demonstration silicon was chosen as the milling substrate in order to minimize charging and ensure high lithographic quality. A droplet of a UV-curable, transparent polymer (Ormocomp) is dispensed on a glass substrate, which was previously treated with hexamethyldisilazane (HMDS), and the Si master mold is cast into it (Fig. 1b). After material cross-linking under UV radiation, the mold and substrate are manually demolded, yielding replica #1. The replication step inverts lithography's tonality, which results in extruded patterns (Fig. 1c). Due to the low stiffness of cured Ormocomp, a mold based on these extruded patterns is not ideal for use as the final mold. Therefore, replica #1 undergoes a second replication step that restores the original lithography tonality (Fig. 1c, d). Prior to replication, the surface of replica #1 is vapor-coated in a vacuum oven at 90 °C with 1H,1H,2H,2H-perfluorodecyl-trichlorosilane, which serves as an anti-adhesion layer. Replica #2 is obtained using the same process and materials as used for replica #1 (Fig. 1c).

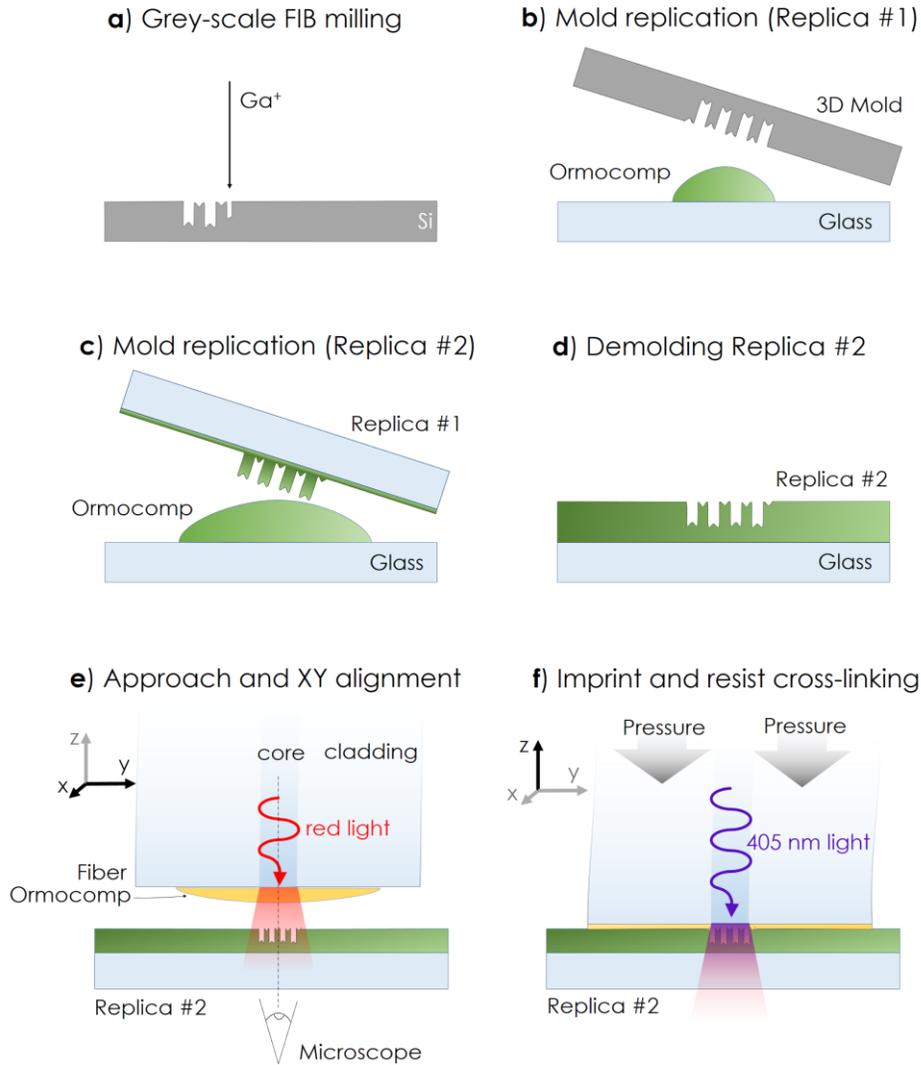

**Figure 1. Process for fabricating a 3D nanostructure on the facet of an optical fiber by NIL**. **a**. A master mold is derived by grayscale FIB milling in silicon. **b**. The master mold is replicated onto glass with Ormocomp, yielding replica #1. **c**. Replica #1 is cast following the same process as in 'b' to obtain replica #2 with the same lithography tone as the master mold. **d**. Replica #2 is demolded and coated with a fluorine-based anti-adhesion layer. **e**. The fiber imprinting step is performed using a custom-made setup and monitored from beneath by an optical microscope. Alignment between the fiber core and structure in the mold is guided by red laser light coupled inside the fiber. **f**. A force is applied to squeeze the resist and allow pattern filling. The resist is cured using 405 nm laser light, which is coupled inside the optical fiber. Demolding is achieved by simply pulling the fiber back.

For nanoimprint onto the facet of an optical fiber, a customized setup was assembled. The transparent mold (replica #2) lies on the XY mechanical stage of an inverted microscope that is
3

focused on the imprint plane and is used to perform the alignment between the structure in the mold and the fiber core (Fig. 1e). The tip of the single-mode (SM) fiber is cleaved flat and immersed into a solution of Ormocomp and dichloromethane at 10 wt%. The amount of Ormocomp that wets the facet of the fiber is determined by the resist dilution, which also enables precise control of the residual layer after imprint. The fiber is loaded into a custom-designed holder, mounted on an XYZ vertical piezo-stage, and approaches the surface of the mold perpendicularly. A red laser is coupled into the fiber and marks the position of the core during alignment (Fig. 1e). Coaxial alignment between the mold structure and the fiber core is carried out under the microscope. Coarse alignment is performed using the microscope XY stage, while piezo-driven stage movements enable sub-200 nm alignment accuracy. Upon contact and pattern filling, laser light with a wavelength of 405 nm is coupled into the fiber to crosslink the Ormocomp (Fig. 1f). The photo-initiators inside the resist function up to this wavelength and light scattering ensures that crosslinking occurs over the entire fiber facet. Finally, demolding is achieved simply by pulling the fiber upward.

To prove the effectiveness of the process described, we chose to fabricate a convoluted 3D photonic structure that splits light of wavelength 405 nm into four beams of equal intensity upon exiting the fiber. An iterative Fourier transform algorithm (IFTA) is used to calculate the phase mask required to perform the beam splitting operation[14]. The ideal height profile of the photonic structure is determined based on the calculated phase mask and refractive index (RI) contrast between the imprint polymer (Ormocomp, RI=1.5) and air.

The resulting height map has 255 levels and is used as a mask for milling, which we performed by Gallium-FIB using a Zeiss Orion NanoFab multibeam ion microscope[15]. A total ion dose of 3.5 nC/$\mu m^2$ at a current of 50 pA and using an accelerating voltage of 30 kV was implemented. Figure 2a shows a color-scale version of the map used for the milling: red and blue regions receive higher and lower ion doses respectively. For the actual milling, a grayscale bitmap version was imported into a NanoPatterning Visualization Engine (NPVE) software coupled to the microscope. Figure 2b shows a Helium Ion Microscopy (HIM) image of the FIB-milled mold. The nanostructure consists of periodic wave-like features, which have a smooth profile within each elementary cell and sharp height discontinuities at the edges of each cell. The discontinuities correspond to a $2\pi$ shift in the phase mask. The spatial periodicity of the structure is ~1 µm but the lithographic resolution is much higher, enabling both smooth height variation as well as sharp discontinuities, as per design. The 5.5×5.5 µm² footprint of the structure is designed to intersect most of the light that exits the core of the SM optical fiber.

Tilted-view scanning electron microscopy (SEM) images of the 3D splitter imprinted on the core of the optical fiber after demolding are shown in Fig. 2c and d. SEM images do not reveal any major deformation of the imprinted geometry, despite the sequence of three replications performed. This clearly demonstrates the capability of our approach to imprint 3D structures on a fiber end with very good fidelity.

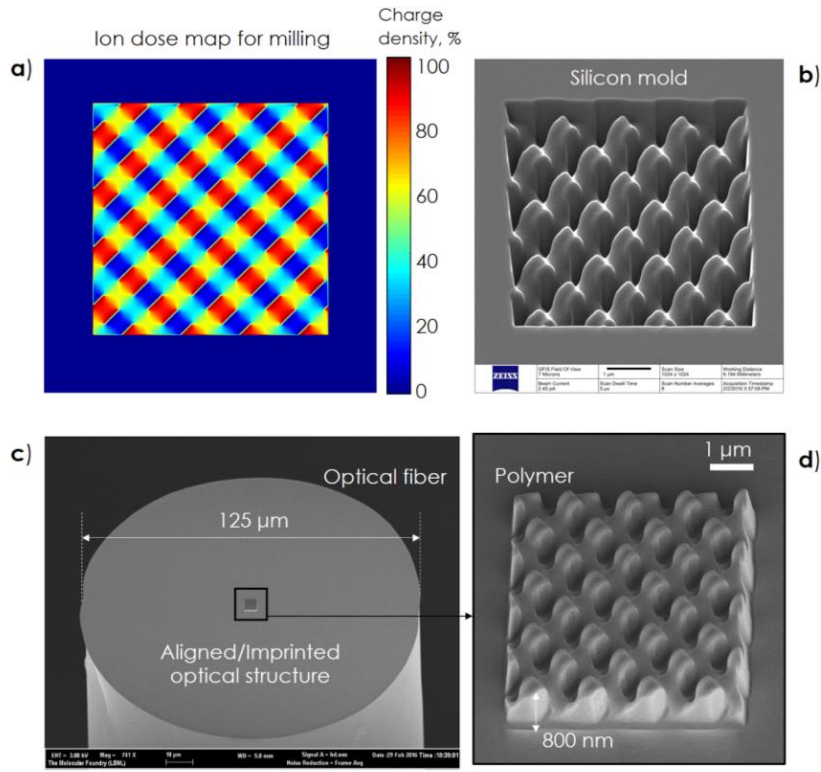

**Figure 2. a**. Color version of ion dose map (255 levels) used for grayscale milling. Map is derived using a proportional relation for the phase mask calculated by IFTA. **b**. HIM image of the milled master mold in silicon. **c**. SEM tilted-view of an imprinted fiber. The 3D structure is aligned to the fiber core to a very good accuracy. **d**. Close-up SEM image of the imprinted 3D structure.

**Optical characterization**

To demonstrate the functionality of the device, light from a 405 nm laser diode is coupled into the imprinted fiber. A light diffuser is placed 10 cm away from the tip of the fiber, perpendicular to its axis. Laser light exiting the fiber creates a far-field pattern on the diffuser screen. The image of the far-field pattern is recorded by a camera that is placed on the other side of the diffuser. Figure 3a shows the measurement of the far-field image obtained using a bare SM fiber prior to imprinting. Figure 3b illustrates the IFTA simulation result for the far-field projection generated by the integrated beam splitter. According to the simulation, light is mostly concentrated in the first diffraction order, appearing as four spots of intensity U, while zero and higher orders have negligible intensity. Figure 3c shows the far-field measurement from the fiber with the imprinted splitter. The simulation and experimental results are in excellent agreement both in terms of beam diameter and field distribution. Higher diffraction orders are in fact also observed in the experiment, probably caused by smoothing of the photonic structure as a result of the milling and replication process. Since an opaque screen cannot be used for accurate power measurement of each spot intensity due to the different ratios of light scattered by the diffuser for varying angles, a photodiode is used for the power measurement instead. The intensities at the four spots are found to vary between 0.94·U and 1.1·U. This result confirms the good quality of the imprinted structure, validating this process to fabricate complex 3D structures on the facet of an optical fiber.

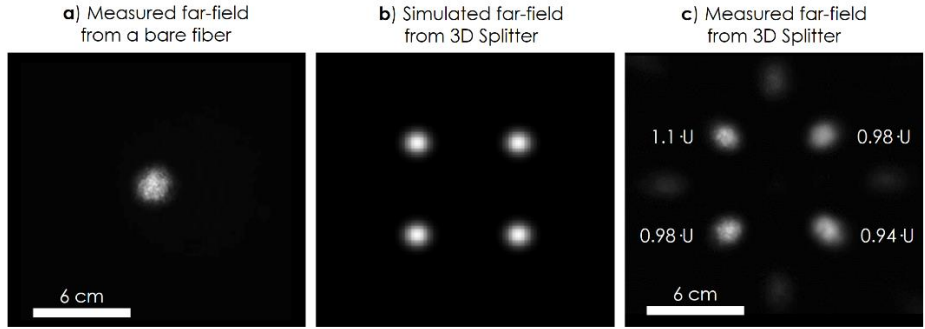

**Figure 3**. **a**. Photograph of the far-field image obtained from a bare fiber prior to imprinting. **b**. IFTA simulated far-field light intensity generated by the phase mask designed. The spots are of equal intensity, denoted here by the arbitrary intensity value "U". **c**. Photograph of the far-field light distribution produced by the 3D splitter imprinted onto a fiber. The distance between the fiber and the image plane is 10 cm. The ratio between the measured intensities and U is reported next to each spot.

**Conclusion**

In this work, NIL-enabled fabrication of a 3D pattern on the facet of an optical fiber for beam splitting was reported. Highly convoluted, sub-wavelength features were fabricated on a fiber facet with good imprint fidelity and alignment accuracy. The imprinted structure enabled precise manipulation of the wavefront of light and generated the desired light distribution in the far-field, as defined by the particular structure designed. Optical characterization of the imprinted fiber splitter demonstrates the excellent performance of the device, with experimental results in good agreement with predictions from simulations. To the best of our knowledge, this is the first example of a complex, 3D NIL process performed on an optical fiber that presents such a high lithographic accuracy.

The process presented can be applied to the fabrication of a variety of other 3D photonic devices to be imprinted directly onto the facet of an optical fiber in a single step and with a high throughput. This approach has the potential to generate a new class of inexpensive fiber optic probes and sensors with applications in laser machining, biology, medicine, and integrated optics.

**Acknowledgment**

This work is supported by the U.S. Department of Energy, Office of Science, Basic Energy Sciences, under Award Number DE-C0013109. Work at the Molecular Foundry was supported by the Office of Science, Office of Basic Energy Sciences, of the U.S. Department of Energy under contract no. DE-AC02- 05CH11231. The Zeiss ORION NanoFab microscope is located at the Biomolecular Nanotechnology Center/QB3-Berkeley and was funded by an NSF grant from the Major Research Instrumentation Program (NSF Award DMR-1338139).